\documentclass{elsart3p}

\usepackage{amssymb}
\usepackage{amsmath}
\usepackage{graphicx}

\begin{document}

\begin{frontmatter}


\title{Dirac Fermions in Graphite: the State of Art}
\author{Igor A. Luk'yanchuk $^{a,}$$^b$\corauthref{cor1}}, \author{Yakov
Kopelevich$^{c}$} and \author{Mimoun El Marssi $^{a}$}
\corauth[cor1]{Corresponding author, E-mail: lukyanc@ferroix.net}
\address{$^a$ University of Picardie Jules Verne, Laboratory of Condensed Matter Physics,
Amiens, 80039, France}
\address{$^b$ L. D. Landau Institute for Theoretical Physics, Moscow, Russia}
\address{$^c$ Instituto de F\'{\i}sica "Gleb Wataghin", Universidade Estadual de Campinas,
Unicamp 13083-970, Campinas, Sao Paulo, Brazil}

\begin{abstract}
Macroscopic concentration of massless charge carriers with linear
conic spectrum - Dirac Fermions (DF) was shown in 2004 to exist in
Highly Oriented Pyrolytic Graphite (HOPG) and governs its electronic
properties. These carriers can have the same nature as DF observed
in graphite monolayer (Graphene) and let to view HOPG as
superposition of 2D carbon layers, almost independent
electronically.  We overview here the recent experimental evidences
of 2D Dirac Fermions in Graphite and their similarity with carriers
in Graphene.
\end{abstract}

\begin{keyword}
Graphite, Graphene, Dirac Fermions, Quantum Hall Effect
\PACS 81.05.Uw \sep 71.20.-b
\end{keyword}
\end{frontmatter}

% main text
\section{Do Dirac Fermions present in graphite} \label{}

Recent synthesis of graphite monolayer (Graphene)
\cite{Novoselov_2005,Zhang_2005} activated study of several
challenging fundamental problems in Solid State Physics related with
electronic properties of semimetals i.e. systems in which several
branches of the spectrum cross close to Fermi level. It was
recognized since a long time \cite{Brandt} and confirmed
experimentally recently \cite{Novoselov_2005,Zhang_2005} that
peculiarity of electrons in graphite monolayer consists in the
particular linear spectrum of charge carriers expanding in vicinity
of the corner point K of the hexagonal Brillouin zone (BZ), similar
to the conical dispersion of massless Dirac Fermions (DF) in 2D
Quantum Electrodynamics.

\begin{figure}[!bh]
\centering
\includegraphics [width=5.5cm] {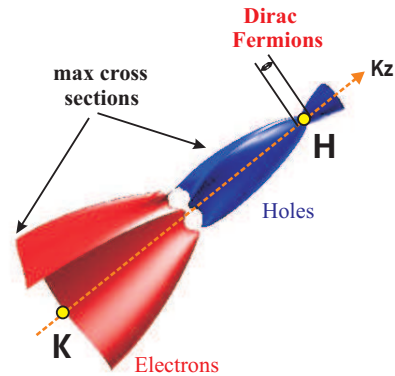}
\caption{According to Slonzewski, Weiss and McClure (SWM) band model
\cite{model}, Fermi surface (FS) in graphite is expanded along $k_z$
edge K-H of the hexagonal Brillouin zone (BZ) and consists of
electron and hole pockets. The maximal cross section of pockets,
seen in quantum-oscillation experiments have the massive dispersion
of charge carriers whereas Dirac Fermions (DF) are concentrated in
the vicinity of H-point with very small phase volume. } \label{FS}
\end{figure}

Note however that according to the tight-binding calculations
\cite{Partoens07}, the linear Dirac spectrum is the specific
property of Graphene. Already the bilayer graphite films have the
electronic dispersion that is presented as junction of up- and down
curved parabolas. Although the Dirac-like branches can also appear
in the n-layered ($n>2$) graphite, their relative phase volume is
negligible and no manifestation of DF should be observed.

At large $n$ the multilayered system recovers the properties of bulk
graphite and the classical Slonzewski, Weiss and McClure (SWM) band
model \cite{model} should be applicable. According to this model the
Fermi surface (FS) of graphite is expanded around z-directed K-H
edge of 3D BZ and have two almost compensated electron and hole
pockets as shown in Fig. \ref{FS}. The perpendicular Dirac-like
dispersion exists only in the immediate vicinity of the point H and
therefore the relative concentration of DF should be negligibly
small.

\begin{figure}[!b]
\centering
\includegraphics [width=7cm] {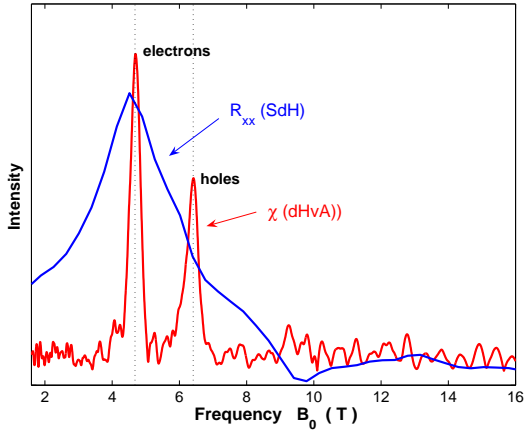}
\caption{Fourier spectrum of de Haas van Alphen (dHvA) and Shubnikov
de Haas quantum oscillations in HOPG reveals two types of carriers:
normal electrons and Dirac holes. The identification of carriers is
done by the phase analysis in \cite{Lukyanchuk_2004}.}
\label{FigSpectr}
\end{figure}

Two majority groups of carriers seen by many researchers in de Haas
van Alphen (dHvA) quantum-oscillation experiments in 50's-60's
\cite{Brandt} (cf. also Fig. \ref{FigSpectr}) were naturally
attributed to the maximal electron- and hole- cross sections of FS
pockets, but no definite conclusion about their spectrum (Dirac or
massive) was done.

In the beginning of 2004, basing on the phase analysis of Shubnikov
de Haas (SdH) and dHvA oscillations in bulk Highly Oriented
Pyrolytic Graphite (HOPG) two of us (IL\&YK ) discovered that one of
these groups corresponds to the macroscopic quantity of  DF
\cite{Lukyanchuk_2004}. Evident discrepancy with SWM 3D band model
concerning the quantity of DF in graphite let to the alternative
view of graphite as the stack of almost independent mono- and multi-
2D carbon layers.

Since then,  several  independent confirmation of existence of DF in
graphite where done. The objective of this communication is to
overview the most important experimental results that let to judge
about the nature of charge carriers in graphite.

\section{Dirac Fermions from quantum oscillation} \label{}

Most of the methods permitting to distinguish between normal
(massive) carriers and DF are based on  Landau Level (LL)
quantization in magnetic field.  In the massive case the equidistant
LLs
\begin{equation} E_n=(e\hbar /m_\perp c)B(n+1/2) \label{En}\end{equation}
 are separated by the gap $E_0=e\hbar B/2m_\perp c$
from $E=0$ whereas in the Dirac-like case the square root dependence
takes place:
\begin{equation} E_n=\pm v \sqrt{2e\hbar B n/c}  \label{Ed} \end{equation}
and the lowest LL is located exactly at $E_0=0$. This, in
particularly, leads to the difference in  Bohr-Sommerfeld
semiclassical quantization:
\begin{equation}
S(E_f)=(n+\gamma )2\pi \hbar \frac{eB}{c}, \label{Bohr}
\end{equation}
for which $\gamma =1/2$ for massive carriers and $\gamma =0$ in the
Dirac case \cite{Mikitik99}. This factor is uniquely related to the
topological Berry phase $\Phi_B=k\pi$ (with $k=2\gamma+1$ (mod 2))
acquired by a fermion, moving around $S(E_f)$ \cite{Mikitik99} and
 is manifested experimentally as the phase in SdH
oscillations of conductivity
\begin{equation}
\Delta \sigma_{xx}(B) \simeq - A(B) \cos[2\pi({\frac{B_0
}{B}}-\gamma+\delta) ]
\end{equation}
or in another quantum oscillations. (FS dependent factor
$|\delta|<1/8$)

Determination of $\gamma$ from SdH and dHvA oscillations was the
base of our phase analysis \cite{Lukyanchuk_2004}. Since, as was
already mentioned,
 two majority groups of carriers are present in graphite  (Fig. \ref{FigSpectr}),
 we first filtered the
oscillating signal from each of them and then measured their phases
separately. Moreover, comparison of SdH and dHvA experiments let us
to determine the sign of carriers. Conclusion was that, the
lower-frequency peak in Fig. \ref{FigSpectr} corresponds to massive
electrons whereas the higher-frequency peak to Dirac holes.
Coincidence of results of SdH and dHvA phase analysis shows that
existence of DF in graphite is the bulk and sample-independent
effect.

Contradiction with band theory is not only in the predicted by SWM
model massive character of both groups of carriers \cite{Mikitik06}
(their maximal cross sections $S_e$, $S_h$ are far from
 Dirac point H, see Fig. \ref{FS}), but also in the inverse order
 of proportional to $S_e$ and $S_h$ oscillation frequencies: it was previously assumed  \cite{Brandt} that
 $S_e>S_h$.

\section{Landau Level spectroscopy}

Alternative way to distinguish between different types of carriers
 is the direct determination  of the energy distribution
of quantized LLs. In case of massive carriers the inter-LL distance
scales linearly with $B$ as in Eq. (\ref{En}) whereas in the
Dirac-like case as $\sqrt{B}$ as in Eq. (\ref{Ed}).

Low temperature (4K) Scanning Tunnelling Spectroscopy (STS)
measurements on the surface of HOPG in fields up to 12 Tesla were
performed to test such LLs energy distribution \cite{Andrei}. It was
discovered that the resulting signal reveals the traces of both
massive and massless LLs, that agrees with  our phase analysis
Unfortunately the natural restriction of
 STS measurements over the sample surface permits also the alternative
interpretation of the origin of DF as due to the surface effects
\cite{Andrei}.

Another measurements of LLs distribution in graphite (both in HOPG
and in natural) that are free from the mentioned above surface
sensitivity  were done using the Far-Infrared (FIR)
magneto-transmission spectroscopy \cite{Orlita_2008} in which the
absorbtion lines correspond to the optical transition between
different LLs. Two series of LLs with normal and Dirac-like
quantization in $B$ were found and attributed to point K and H of 3D
BZ in SWM band model. The density of DF was found to be smaller then
that, extracted from analysis of SdH oscillations
\cite{Lukyanchuk_2004}.

Although authors of \cite{Orlita_2008} agree that the dominant
transitions in the Dirac series of absorbtion lines have their
counterpart in the FIR spectra of graphene \cite{FIRgraphene}, they
believe that the 3D band model is more adequate to describe the
situation because of observation of the weak series of additional
transitions, forbidden in pure 2D case. Note however that in this
case one more series of FIR transitions, corresponded to the maximal
cross section of the hole FS pocket $S_h$ (Fig. \ref{FS}) with
massive LL quantization should be observed.

\section{Angle Resolved Photoemission}

Coexisting of massless DF  with quasiparticles with finite effective
mass in graphite was also proven by  Angle Resolved Photoemission
Spectroscopy (ARPES) \cite{Zhou}. This method is the most direct and
unique tool to measure the electronic structure with both energy and
momentum information. The electronic spectrum was scanned in the
inverse space of BZ in vicinity of points K and H in the
$\textbf{k}\perp \textbf{k}_z$ direction. The electronic dispersion
was shown does correspond to the massive carriers in K-point and to
DF in H-point as was predicted by SWM band model.

Although ARPES measurements convincingly demonstrate the presence of
DF in graphite, no further information about their quantity and
macroscopic properties can be extracted. To know their concentration
and to check the validity of the assumed in \cite{Zhou} 3D band
model the further  scans in $\textbf{k}\perp \textbf{k}_z$ direction
along whole K-H edge are required.

\section{2D Transport and Quantum Hall Effect}

To give more insight in favor of 2D independent layered structure of
graphite we refer on another group of experiments. Note first that
because of the  extremely high out- to in- plane transport
anisotropy in HOPG: $\rho_{out}/\rho_{in} \sim 5\cdot 10^4$
\cite{Kopelevich_2003} (instead of $300$ as was measured for some
graphite crystals \cite{Brandt}) it is unlikely that out-of-plane
transport can be explained within the inter-plane hopping of SWM
model. Moreover, the characteristic property of 2D conducting
systems - the Quantum Hall Effect (QHE) has been discovered in 2003
in HOPG \cite{Kopelevich_2003}.

Careful analysis of  QHE staircase in HOPG \cite{Lukyanchuk_2006}
shows that it contains the precursors of both QHE discovered in
graphene \cite{Novoselov_2005,Zhang_2005} and in bilayer graphite
\cite{Novoselov_2006}, having the semi-integer (Dirac)
\cite{Gusynin05,Peres06} and integer (massive) \cite{McCann_2006} LL
quantization correspondingly. Order and appearance of normal and
Dirac steps in HOPG depends on the sample \cite{Kopelevich_2007} and
sometimes coincides with QHE staircase observed in few-layer
graphite sample \cite{Novoselov_2004}. Such behavior confirm the
hypothesis that bulk graphite is a system randomly composed of
single- and few- carbon layers.

\section{Raman fingerprint of Graphene in graphite}

Next, we present the results of our recent micro-Raman testing of
the best (in 2D transport sense) HOPG samples. (The details will be
published elsewhere.) Raman study of the  second order peak "2D",
which is provided by specific to Dirac spectrum  double-resonance
electron transitions \cite{Thomsen}, is known to be the best tool to
distinguish the mono- and multilayered graphite films
\cite{Ferrari}.

\begin{figure}[!t]
\centering
\includegraphics [width=7cm] {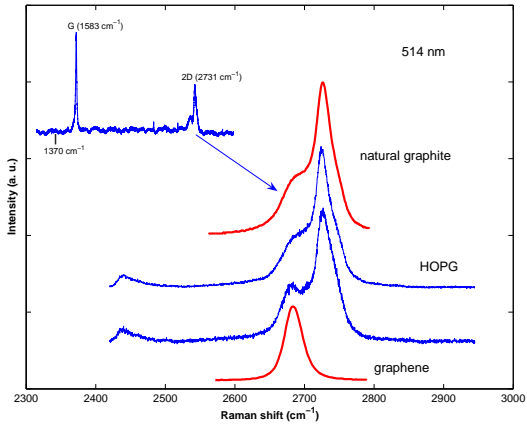}
\caption{Double resonance 2D micro-Raman band of HOPG graphite (two
blue plots in the middle) demonstrates the superposition of the
features typical for bulk natural graphite (upper plot with
left-shoulder profile) and for the monolayer graphene (lower
single-peak plot) \cite{Ferrari}. The relative weight of these
features depends on the position of the laser spot on the surface of
HOPG. The typical Raman spectrum of graphite is shown in the
upper-left corner } \label{FigRaman}
\end{figure}

Shown in Fig. \ref{FigRaman} the upper left-shoulder-peak "2D" in
bulk graphite   is indeed very different from lower, specific only
to monolayer the single-Lorentzian  peak of graphene \cite{Ferrari}.
Two middle plots correspond to our micro-Raman testing of HOPG with
different position of few-micron laser spot on the surface.

Surprisingly we found that depending on the spot position, the
classical graphite peak can demonstrate the growing fingerprint
feature of graphene. Because of several microns laser penetration
length, such effect  can be explain only by existence of macroscopic
number of independent graphene layers in bulk of HOPG. Since this
conclusion is coherent with other considered above experimental
facts we can express even more general hypothesis that left shoulder
in peak "2D" in graphite is provided in total by the presence of
graphene monolayers in the bulk.

\section{Conclusion}

Retrospective overview of the experimental data let to believe that
macroscopic concentration of DF does exist in the bulk of graphite.
Since this conclusion contradicts to classical SWM band model,
further experimental and theoretical efforts should be done to
clarify their nature. It is of primary importance to understand,
whether the modified band theory should be constructed to explain
the existence of DF  or  indeed, graphite can be viewed as the
system of independent mono- and multi carbon layers as proposed in
the present article.

This work was supported by French-Brazilian exchange program
CAPES-COFECUB, by the ANR grant LoMaCoQuP (France) by agencies CNPq,
FAPESP (Brazil) and by project ROBOCON of EU F7 IRSES program.

\end{document}